\documentstyle[twocolumn,aps,psfig]{revtex}
\begin{document} 
\draft
\title{Spin-dependent electrical transport in ion-beam sputter deposited
 Fe-Cr multilayers}
\author{A. K. Majumdar\cite{akm} }
\address{Department of Physics, Indian Institute of Technology, Kanpur-208016, India}
\address{Department of Physics, University of Florida, Gainesville, Florida 32611}
\author{A. F. Hebard}
\address{Department of Physics, University of Florida, Gainesville, Florida 32611}
\author{Avinash Singh}
\address{Department of Physics, Indian Institute of Technology, Kanpur-208016, India}
\author{D. Temple}
\address{MCNC, Electronics Technologies Division, Research Triangle Park,
North Carolina 27709}
\maketitle
\begin{abstract} 
The temperature dependence of the electrical resistivity and 
magnetoresistance of Xe-ion beam sputtered Fe-Cr
multilayers has been investigated.
The electrical resistivity between 5 and 300 K in the 
fully ferromagnetic state, obtained by applying a field beyond the 
saturation field ($H_{\rm sat}$) necessary for the
antiferromagnetic(AF)-ferromagnetic(FM)
field-induced transition, shows evidence of spin-disorder 
resistivity as in crystalline Fe and an s-d scattering  
contribution (as in 3d metals and alloys). The sublattice magnetization 
$m(T)$ in these multilayers has been calculated in terms of the planar and 
interlayer exchange energies. The additional spin-dependent scattering 
$\Delta\rho(T)=\rho(T,H=0)_{AF}-\rho(T,H=H_{\rm sat})_{FM}$ in the AF state  
over a wide range of temperature is found to be proportional to the  
sublattice magnetization, both $\Delta\rho(T)$ and $m(T)$ 
reducing along with   
the antiferromagnetic fraction. At intermediate fields, 
the spin-dependent part of the electrical resistivity  $(\rho_{s}(T))$ 
fits well to the power law $\rho_{s}(T) = b - cT^{\alpha}$ where $c$ is a constant
and $b$ and $\alpha$ 
are functions of $H$. At low fields $\alpha \simeq 2$ and the intercept  
$b$ decreases with $H$ much the same way as the decrease of 
$\Delta \rho (T)$ with $T$.
A phase diagram ($T$ vs $H_{\rm sat}$) is 
obtained for the field-induced 
AF to FM transition. Comparisons are made between the  
present investigation and similar studies using dc magnetron  
sputtered and molecular beam epitaxy (MBE) grown Fe-Cr multilayers. 
\end{abstract} 

\section{Introduction}
As one of the very few lattice matched transition metal pairs 
one of which is ferromagnetic, Fe-Cr multilayers offer excellent
opportunities for investigating the exchange coupling of Fe layers
through an antiferromagnetic Cr spacer layer giving rise to the 
so-called  Giant-Magnetoresistance (GMR). Applications as 
magnetic-field sensors, especially in reading information,
sensing position/speed of moving parts, etc. have triggered intense 
research activity in these multilayers. GMR sensors are not only
very sensitive  but they can be made very small in size. 
For practical purposes not only does one need large GMR but also small
saturation fields. 

The aim of the present work is to study the temperature dependence
of electrical resistivity and magnetoresistance in GMR multilayer
stacks prepared by ion-beam sputter deposition. 
Typical multilayers reported here comprise 30 repeat layers of 
[Fe(20\AA)/Cr(10\AA)] that have been deposited by ion-beam sputter 
deposition onto Si substrates with Xenon ions at 900 V
and a beam current of 20 mA.

GMR in multilayers can be understood in terms of some simple ideas 
as follows. In zero magnetic field the ferromagnetic Fe layers
are coupled antiferromagnetically through the Cr spacer layer giving 
rise to  a high electrical resistance. 
This antiferromagnetic coupling is ascribed to the indirect exchange
interaction between the Fe layers through the oscillatory RKKY 
interaction mediated by the conduction electrons.
The above antiferromagnetic coupling between Fe layers was established 
by means of light scattering from spin waves.\cite{one}
As the external field increases, the spins in different Fe layers
align in the direction of the field producing a completely
ferromagnetic alignment beyond a saturation field $H_{\rm sat}$,
reducing the resistance. Thus we have a negative magnetoresistance (MR).

Magnetoresistance is defined (in text-book fashion) by
\begin{equation}
{\rm MR}=\frac{\rho(H,T)-\rho(0,T)}{\rho(0,T)}\times 100 \% \; .
\end{equation}
It is found in dc magnetron sputtered Fe-Cr superlattices\cite{two}
that GMR oscillates as a function of Cr spacer thickness with
3 gradually decreasing peaks centered at 11, 27 and 42 \AA \ Cr for Fe thickness
of 32 \AA. Also the first antiferromagnetic region occurs between 6
and 11 \AA \ Cr for Fe thickness lying between 15 and 40 \AA. 
We have therefore chosen the Cr thickness around 10 \AA \ corresponding
to the strongest peak of the antiferromagnetic coupling between the
Fe layers (and hence the highest GMR).

The basic qualitative features of GMR can be understood even in 
terms of bulk scattering only if the mean free path (mfp) of electrons
within the layers is much larger than the layer thickness.
If the mfp of the electrons is larger than the Cr spacer thickness,
the electrons can feel the relative orientation of the magnetization
of the successive layers. However, this interplay between the 
successive magnetic layers disappears and the GMR vanishes if the mfp
is less than the Cr layer thickness.

GMR is attributed to the spin-dependent conduction properties of 
ferromagnetic metals. In a ferromagnetic metal or alloy the electrical
conduction  takes place through independent channels by spin-up
(called majority) and spin-down (called minority) electrons.
This is the two-current model of Fert and Campbell\cite{three}
whose physical basis is the dominance of the spin-conserving 
scattering and the weakness of the spin-flip collision, at least 
at low temperatures. In this picture all electrons of a given spin 
(up or down) with s or d or hybridized  character are grouped together 
to form majority (up) or minority (down) bands.
If one takes into account the details of the band structure of Fe
(a weak ferromagnet) and the simple Drude conductivity formula 
for each band, it is easily shown\cite{four} that the majority band
has a much higher conductivity than the minority band. As a result, 
in the ferromagnetic alignment brought about by the saturation field
($H_{\rm sat}$), there is hardly any scattering for the majority 
band electrons since they remain majority in all the Fe layers.
On the other hand, the minority band electrons get scattered within
every Fe layer. Hence there is a short-circuiting effect, so to say, 
and the resistance ($\rho_{\rm FM}$) drops in the ferromagnetic 
alignment. However, in the antiferromagnetic configuration (zero-field)
both the majority and minority band electrons are scattered in successive
layers, and there is no short-circuiting effect. Therefore the resistance
($\rho_{\rm AF}$) remains relatively high. A very simple calculation 
based on the two-current model shows that the GMR at low temperatures
is given by (considering only bulk scattering) 
\begin{eqnarray}
{\rm GMR} &=& \frac{\rho(H=H_{\rm sat})-\rho(0)}{\rho(0)}
= \frac{\rho_{\rm FM} - \rho_{\rm AF}}{\rho_{\rm AF}}
\nonumber \\
&=& - \left ( 
\frac{ \rho_\downarrow /\rho_\uparrow - 1 }
{ \rho_\downarrow /\rho_\uparrow + 1 }
\right ) ^2 \; ,
\end{eqnarray}
where $\rho_\downarrow $ and $\rho_\uparrow $ are the resistivities
of the minority and majority carriers, respectively.
It turns out that interface scattering from  imperfect interfaces,
defects and impurities in Fe-Cr multilayers is also spin-dependent,
and gives rise to  the GMR. As a matter of fact, it is the imbalance 
between the resistivities of the two bands which is responsible
for the GMR for bulk, interface and spin-flip scattering.
The subject of GMR has been reviewed very well 
in a recent book.\cite{five}

Considerable work has been reported on electrical transport 
and magnetic properties of Fe-Cr multilayers prepared by 
sputtering\cite{two,six} or Molecular Beam Epitaxy 
(MBE).\cite{six,seven,eight} The temperature dependence of the electrical
resistivity in [Fe(30\AA)/Cr(10 - 50\AA)]$\times 10$ 
multilayers, prepared by sputtering and MBE, was interpreted
by Almeida {\em et al.}\cite{six}
in terms of phonon-assisted s-s and s-d scattering in the temperature
range of 15-300 K and in a saturation magnetic field of 7.5 kOe.

In the antiferromagnetic Fe-Cr superlattice, made by
dc magnetron sputtering, the temperature-dependent 
magnetoresistance 
(defined as 
$ \rho_{\rm M} (T) = \rho(T,H=0) - \rho(T,H=H_{\rm sat})$ )
was found by Mattson {\em et al.}\cite{two} to follow the equation for $T < 100 K $
\begin{eqnarray}
\rho_{\rm M} (T) &=& \rho_{\rm M} (T=0) +
\Delta \rho_{\rm M} (T) \nonumber \\
&=& \rho_{\rm M} (T=0) -a T^2  \; ,
\end{eqnarray}
where $a$ is a constant of proportionality. This behaviour
was explained in terms of thermal excitation of magnons whose occupation
number ($n$) $\propto T^2$ (at low temperatures) for anisotropic materials\cite{nine}
and assuming $\Delta \rho_{\rm M} (T) \sim n $.

In [Fe(12\AA)/Cr(12\AA)]$\times 10$ multilayers, prepared by MBE on
MgO(100) substrates, several interesting observations were made by
Aliev {\em et al.}\cite{eight} 
Among them are: \\
(a) The isothermal magnetoresistance, defined as $[\rho(0)-\rho(H)]$,
is proportional to $H$ in the parallel (magnetic field in the plane of the multilayers)
orientation and to $H^2$ in the perpendicular case. \\
(b)A $T$ - $H_{\rm sat}$ phase diagram was obtained, which clearly indicated the 
transition between AF and FM states. \\
(c) As opposed to the work of Mattson {\em et al.}\cite{two} 
(which looked at the difference between the resistivity in the ideally 
antiferromagnetic ($H=0$) and ferromagnetic ($H=H_{\rm sat}$) alignments)
here\cite{eight} the spin-dependent part of the electrical resistivity, defined as 
$\rho_s (T) = \rho(T,H) - \rho(T,H>H_{\rm sat})$ for fields $H\ge 0 $
(where one has both ferromagnetic and antiferromagnetic fractions),
is found to vary in a wide range of temperature below 100 K as
\begin{eqnarray}
\rho_s (T) &=& \rho_s (T=0) +\Delta \rho_s (T) \nonumber \\
&=& b - c\;  T^ \alpha \; ,
\end{eqnarray}
where $b =  \rho_s (T=0)$ and the temperature exponent $\alpha$ are functions of the magnetic field $H$,
and $c$ is a constant of proportionality.
The constant $\alpha$ was found\cite{eight} to be 
$\simeq 1.7$ for $H\simeq 0$, 
$\simeq 2.0$ for $H < 0.5 H_{\rm sat}$, and
$\simeq 1$ for $H\simeq H_{\rm sat}$.
This is in contrast to the value of $\alpha = 2$ for $H=0$ 
(purely antiferromagnetic) in the work of Mattson {\em et al.}\cite{two}
(Eq. (3) above). \\
(d) $\Delta \rho_s (T)$ was found to vary linearly with temperature from 20 mK to about 1.5 K 
``which could be due to electron scattering on critical thermal
spin fluctuations''.\cite{eight} 

All the work summarized above is mostly on MBE grown Fe-Cr multilayers 
and those made by dc magnetron sputtering. Although the present work is on
Fe-Cr multilayers
prepared by ion-beam sputter deposition, we do not believe that the underlying
physics depends in any significant way on differences in these deposition techniques.

\section{Experimental}

Fe-Cr multilayers were prepared by ion-beam-sputtering technique and
characterized by Transmission Electron Microscopy (TEM), Atomic Force
Microscopy (AFM), Auger Electron Spectroscopy (AES), X-ray Photoelectron
Spectroscopy (XPS), as well as resistivity, magnetic hysteresis loop and
magnetotransport measurements.  The film deposition procedure and film
properties, including chemical composition, surface morphology, resistivity,
saturation magnetization, coercive field, and magnetoresistance ratio, are
discussed in detail in Ref. [10].  Values of GMR ratios of the films are
comparable to  values measured for polycrystalline Fe-Cr films deposited by
the more conventional rf sputtering technique.\cite{Parkin} We have
deposited the following Fe-Cr multilayer combinations :
\\ Si/Cr(50\AA)/[Fe(20\AA)/Cr(t(\AA))]$\times$30/Cr(50-t (\AA)), where $t$ was
varied from 8 to 14\AA; this range surrounds the first 
antiferromagnetic maximum in the Fe-Cr multilayer system. The deposition 
rates varied from 5 to 30 \AA/min depending on the primary ion beam
energy, the type of ions, and the target material. The films were deposited at 
room temerature. The effects of variations of the primary ion beam energy
 and the type of ions on GMR values were examined;
the investigated primary ion energy range was 700-1200 eV for Ar ions and
900-1200 eV for Xe ions. It was demonstrated that the GMR ratio is 
greater for films deposited using Xe ions than for films deposited 
using Ar ions, and that for both types of ions the GMR ratio increases 
as the primary ion beam energy decreases. In this investigation we 
report the work on Fe-Cr multilayers
of typical structure [Fe(20\AA)/Cr(10\AA)]$\times$30 layers grown on
Si substrates using Xe ion at 900 V and a beam current of 20 mA.

The temperature dependence of the resistance between 5 and 300 K of the  
Fe-Cr multilayers was measured in zero as well as in some applied  
magnetic fields using the standard four-probe dc technique and a  
magnetic field (0-5.5T) provided by a Quantum Design SQUID magnetometer 
(MPMS). Both the transport current and the applied field were in the plane of 
the film with the current parallel to the field. For measurements in  
magnetic fields perpendicular to the film plane we used a Quantum Design 
 Physical Property Measurement System (PPMS). We used the same MPMS to  
 measure the magnetization of the Fe-Cr multilayers as a function of  
 external fields at temperatures down to 2 K. 
 
\section {Results and Discussion}
Figure 1 shows the magnetoresistance vs. external field $H$ (kOe)
for a typical Xe-ion sputtered Fe-Cr multilayer 
sample at 10 and 300 K. The MR becomes constant at a saturation field
$H_{\rm sat}$ (vertical arrow) around 
13 kOe with typical values of 21\% at 10 K. These values compare 
favourably with 30\% and 40\% obtained in DC magnetron sputtered 
and MBE grown samples, respectively.
Hysteresis in the MR was negligible
as the magnetic field was swept from 
$(0\rightarrow 20\rightarrow 0\rightarrow -20\rightarrow 0)$ kOe.

\subsection{Temperature dependence of the electrical resistivity}

Figure 2 shows the electrical resistivity $(\rho)$ vs. temperature 
$(T)$ for sample 1 at several values of the external magnetic field
from 0 to 12 kOe. The saturation field ($H_{\rm sat}$)
has a weak temperature dependence, namely, it decreases with 
increasing temperature. This is clear from the $\rho(T)$ curves, say,
at 10 and 12 kOe. They are closer to each other at higher temperatures. 

To interpret the temperature dependence of the electrical resistivity,
$\rho(T)$, at an intermediate field between $H=0$ and $H=H_{\rm sat}$
is not simple. In this magnetic field region
the material has both antiferromagnetic and 
ferromagnetic  fractions. Instead, for $H\ge H_{\rm sat}$, the alignment
of the spins of each Fe layer is parallel to the direction of the 
external field giving rise to a fully ferromagnetic state. 
It is well known in crystalline bulk 3d metals and alloys that the
electron-phonon scattering contribution to the electrical
resistivity from $\rho_{sd}$ dominates over $\rho_{ss}$ due to
the overlap of the s and d bands at the Fermi level.
Specifically, for Fe the density of states of the 3d$^\uparrow$
majority band at the Fermi level is rather large compared to those
of the s bands. The resistivity, $\rho_{sd}$, is given by the Bloch-Wilson 
formula.\cite{ten} The ``spin-disorder resistivity" coming from 
the electron-magnon (spin-wave contribution) scattering is well
described\cite{eleven} by a relatively small term varying
as $T^2$ in ferromagnets like Fe, Co, and Ni. 
Putting all these contributions together along with the residual
resistivity $\rho_0$ one can write, assuming Mathiessen's rule
\begin{eqnarray} 
\rho(T,H=H_{\rm sat})&=&\rho_0 + A \left (\frac{T}
{\Theta_{\rm D}}\right )^3
\int _0 ^{\Theta_{\rm D}/T}
\frac{z^3 dz}{(e^z -1)(1-e^{-z})} \nonumber \\
&+& BT^2 \; ,
\end{eqnarray}
where the second term is the Bloch-Wilson contribution 
$\rho_{sd}$ and the third term is the small 
electron-magnon contribution increasing with temperature due to 
thermal excitation of magnons.

Taking the Debye temperature $\Theta_{\rm D}=420$ K for Fe-Cr multilayers\cite{six},
we have fitted the data for five samples at their respective 
$H_{\rm sat}$ to Eq. (5) using a 3-parameter least-squares fit program
which also evaluated the
integral numerically at each iteration.
Excellent fits were obtained for all the samples 
with correlation coefficients of 0.999995
and values of the normalized $\chi^2$ 
consistent with the number of degrees of freedom and error estimates.
Figure 3 shows $\rho$ vs. $T$ data (points) for three samples from
5 to 300 K at their respective $H_{\rm sat}$. 
The solid lines are the best-fit curves 
to Eq. (5). It is found that the value of the coefficient of the 
magnetic scattering term, $B$, averaged over all the 5 samples is 
$(4\pm 1)\times 10^{-5}\; \mu \Omega$ cm K$^{-2}$
compared to $1.5\times 10^{-5}\; \mu \Omega$ cm K$^{-2}$ in bulk
ferromagnets (Fe, Co, Ni). This higher value of $B$ may be related to the fact
that the resistivity 
of these Fe-Cr multilayers at 300 K is about five times larger than 
that of bulk iron.

The fits of the data of Fig. 3 to Eq. (5) without the magnetic 
$(BT^2)$ term are distinctly inferior to those with the magnetic term.
The values of $\chi^2$ are typically 6 times larger and the correlation
coefficients poorer for the fits without the magnetic term.
The deviation of the actual data from the best-fit values (residuals)
is plotted
in Fig. 4 as a function of temperature for sample 2 for both the fits. 
The deviation is much less ($< 0.1\; \mu\Omega$ cm in 
$40\;\mu\Omega$ cm) and more random for the fits with the magnetic term 
than without it. Addition of a Bloch-Gr\"{u}neissen (BG) term ($\rho_{ss}$),
which has a
$T^5$ dependence,
or replacing the Bloch-Wilson term by the BG term makes
the fit much worse. 

\subsection{Temperature dependence of the magnetoresistance}

At $H=0$ these multilayers are ideally in an antiferromagnetic state 
where the neighbouring ferromagnetic Fe layers are all 
antiferromagnetically coupled resulting in a higher resistivity.
Actually there may be pinholes through the Cr spacer layer directly 
coupling the Fe layers ferromagnetically instead.
Let us define an antiferromagnetic fraction (AFF) as
\begin{equation}
{\rm AFF}(H)=(1- M(H)/M_s)\times 100 \% \; ,
\end{equation}
where $M_s$ is the magnetization measured at $H=H_{\rm sat}$ 
and $M(H)$ is the magnetization when the field is
reduced from saturation to $H$, all at 5 K.
Our $M(H)$ measurements on these samples show that the AFF is typically
80\% at $H=0$. As $H$ is increased the Fe layers gradually 
turn their magnetization in the direction of the external field reducing 
the AFF and hence the resistivity. Finally the AFF reduces to zero
(fully ferromagnetic alignment) and the resistivity and hence the 
GMR saturates at $H=H_{\rm sat}$.

We define $\Delta \rho (T)=\rho(T,H=0)_{\rm AF} -
\rho(T,H=H_{\rm sat})_{\rm FM}$ as the difference in resistivity
at a given temperature $T$ between the AF $(H=0)$ and the 
FM $(H=H_{\rm sat})$ states, both assumed ideal. This $\Delta \rho(T)$
is primarily due to the additional spin-dependent scattering 
(both bulk and interface) in the antiferromagnetic (AF) state.
It is assumed here that the residual resistivity and the interband 
s-d scattering (dominant for 3d metals and alloys) do not depend
strongly on magnetic fields. Figure 5 plots $\Delta \rho(T)$ vs. $T$
data (stars) for samples 1 and 3. Thus the additional spin-dependent 
scattering (resistivity) in the AF state decreases with increasing temperature. 
Just like the magnetic field aligns the spins in different Fe layers 
reducing the AFF (gradually bringing ferromagnetic order in its place)
and produces a negative magnetoresistance, here temperature reduces the
antiferromagnetic order (potentially bringing down the AFF) and hence 
$\Delta \rho(T)$. It is seen from Fig. 5 that $\Delta \rho$ varies
as $T^2$ at low temperatures and is roughly linear at higher temperatures.

Singh {\em et al.}\cite{nine} had worked out the reduction in
antiferromagnetic order due to thermal excitation of 
spin waves in highly anisotropic antiferromagnets 
with weak interlayer coupling between the antiferromagnetic planes.
This theory has been extended in the present case 
where each Fe layer is ferromagnetic but 
coupled antiferromagnetically to the neighbouring Fe layers due to 
the RKKY interaction. 
In terms of the planar and interlayer exchange energies 
$J_p$ and $J_z$, respectively, the sublattice magnetization 
$m(T)$ at temperature $T$ is given in the Appendix (Eq. (13)).
Assuming $\Delta \rho(T) \sim m(T)$, we have the relation
\begin{equation}
\frac{\Delta \rho(T)}{ \Delta \rho(0)}
=
1 - \frac{1}{\pi^2} \frac{T}{J_p} 
\int_{0} ^{\frac{\pi}{2}} dq_z \;
\ln \left ( \frac{1}
{1-e^{-\frac{J_z}{T}(1-\cos ^2 q_z )^{1/2}} }
\right ) \; .
\end{equation}
where $m(0)=1$.
The expression on the right differs only insignificantly from the 
corresponding expression for $m(T)/m(0)$ obtained earlier\cite{nine} 
for the anisotropic antiferromagnet
($\cos ^2 q_z$ instead of $\cos q_z$),
where it was shown to fall off as $T^2$ at 
low temperatures ($T \ll J_z$),
crossing over to an approximately linear ($T\ln T$) fall off 
at high temperatures ($T \gg J_z$). We note that this expression is relatively
unchanged when ferromagnetic domains are included.
  
We numerically evaluated the
integral in Eq. (7) and used a three parameter least squares fit 
program to fit the data of Fig. 5. The resulting best-fit curves,
shown by solid lines in Fig. 5, yield values of $\chi^2 $ consistent
with the experimental errors and a correlation coefficient of 0.9999.
Estimates of $J_p$ and $J_z$ are found from the above fits.
They are $(230\pm 20)$ K and $(70 \pm 20)$ K, respectively.
The value for $J_p$ is well below the Curie temperature (1040 K) 
for bulk iron but could be closer to the unknown Curie temperature 
of $20\AA$ thick Fe films sandwiched between Cr spacer layers. 
The value for $J_z$ is however satisfying, close to the recently identified 
glass temperature, $T_g$=140 K, of an antiferromagnetic glassy phase 
that coexists with GMR in similar multilayer films. Irreversibilities in this 
glassy phase have been shown to arise from the same interlayer coupling that 
drives the antiparalllel alignments in GMR.\cite{twelve}

Consistent with the above model are experimental data taken by us
(not shown) and others\cite{seven} in
which the saturation fields are studied as a function of
spacer layer thickness.
For three ion-beam sputter deposited 10-layer samples of 
Fe(20\AA)/Cr($d_{\rm Cr}$) with different
Cr spacer layer thickness ($d_{\rm Cr}$) we have observed that
as $d_{\rm Cr}$ increases from 8
to 12\AA, the saturation field $(H_{\rm sat})$ decreases from 
10 to 5 kOe.
As $J_z$ decreases with increasing 
d$_{\rm Cr}$, smaller external fields are necessary to break the
antiferromagnetic coupling between the Fe layers. 

\subsection{Temperature dependence of the magnetoresistance in 
intermediate fields ($0 \le H \le H_{\rm sat}$) }

Following the work of Aliev {\em et al.}\cite{eight} 
we have fitted our data for samples 
1, 2, and 3 to Eq. (4) with $\rho_s(T)=\rho(T,H) - \rho(T,H_{\rm sat})$. 
The data points along with the least-squares fit curves are shown in 
Fig. 6 for sample 1 for $H$=0,3,4,6, and 8 kOe. 
Excellent fits are obtained for all the samples with values of $\chi^2$
consistent with the experimental error, correlation coefficients $R^2 
\simeq 0.999$ and small errors
in the fitting parameters $b$, $c$ and $\alpha$. The fits are, however, 
better for smaller fields. Figure 7 shows $\alpha$ of Eq. (4) vs. $H/H_{\rm sat}$
for all the three samples. The solid lines are just guides to the eye.
The shape of the curve is rather similar to the results obtained by 
Aliev {\em et al.}\cite{eight}
(summarized at the end of INTRODUCTION and Fig. 3 of Ref. [8]) for MBE grown samples.
However, we find some differences, like $\alpha$ in our work is typically 2 for
$H/H_{\rm sat} < 1/3$, becoming $\simeq 1$ for 
$H/H_{\rm sat} \simeq 2/3$
and decreasing at still higher fields. 
This implies that $\rho_s$  vs. $T$ 
curves (Fig. 6) are quadratic in lower fields and linear around 
$H/H_{\rm sat} \simeq 2/3$ instead of $\simeq 1$ as in the work of
Aliev {\em et al.}\cite{eight}

It is found from Fig. 6 that the intercept $b$ decreases with increasing
applied fields. 
This is simply due to the fact that the antiferromagnetic fraction (AFF)
decreases with increasing field. If we plot $b$ as a function of our
measured values of AFF (\%) for samples 1 and 3 we find, as shown in Fig. 8, that $b$ 
increases with the AFF in a monotonic fashion (both decreasing with $H$).
This is a logical conclusion since as $H \rightarrow 0$, the AFF attains its maximum
value giving the highest resistivity in the ideally AF ground state. 
It is to be noted that 
the decrease of $b$ with $H$ and the decrease of $\Delta \rho$ 
with $T$ (Fig. 5) have a common origin. 
It is the decrease of the AFF brought about by $H$ and $T$, respectively.

\subsection{$T$ vs. $H_{\rm sat}$ phase diagram for the AF-FM transition}

Figure 9 shows the low-field magnetoresistance of sample 4
(argon-ion sputtered) vs. external field $H$ for the parallel
orientation ($H$ in the film plane). It is amply clear that 
the MR $\sim H^2$ in contrast to the findings of 
Aliev {\em et al.}\cite{eight} (Fig. 1(a) of Ref. [8])    
who found that the MR is linear in $H$ for their MBE grown samples.
We found no linear region in the MR vs. $H$ curve even at higher
fields until the saturation field ($H_{\rm sat}$) of 2-3 kOe was
reached. As a matter of fact sample 4 (Fig. 9) reflects an 
S-shaped curve having points of inflection.
However, for the perpendicular orientation 
($H$ perpendicular to the film plane) we find that the MR again goes as 
$H^2$ in agreement with the findings of Aliev {\em et al.}\cite{eight}

The observed $H^2$ dependence at low external fields is, 
in fact, expected from 
simple energy considerations, as argued below.
The antiferromagnetic ground state of the multilayer is 
characterized by the sublattice magnetization 
${\bf m} = ({\bf m}_A - {\bf m}_B)/2$, 
which takes into account the antiparallel orientation of the 
spin polarization in alternating Fe layers A and B. 
The direction of ${\bf m}$ is arbitrary in the ideal 
isotropic situation. 
When a small in-plane magnetic field is applied, 
the sublattice magnetization ${\bf m}$ 
aligns itself perpendicular to the direction of the field.
This is the lowest-energy configuration
as it allows for energy gain in all Fe layers due to twisting 
of spins in the field direction.
If the twist angle is $\theta$, assumed small, then the energy gain is
$m H \sin \theta \approx m H \theta $.
The twisting also costs energy 
$J_z m^2 (1-\cos 2\theta ) \approx 2J_z m^2 \theta^2 $
due to loss of antiferromagnetic exchange energy at the layer interfaces. 
Minimizing the net energy change yields the optimum twist
angle, $\theta(H) = H/4Jm$, as proportional to the field.
Now the reduction in the sublattice magnetization or the 
antiferromagnetic fraction (AFF),
and therefore the decrease in resistivity,  
due to this twist is $m(1-\cos \theta(H))$, 
which goes as $H^2 $ for low fields. 

Figure 10 shows the $T$ vs. $H_{\rm sat}$ phase diagram for sample 4
(argon-ion sputtered, same as that of Fig. 9) and sample 1  
(xenon-ion sputtered). Both are in parallel orientations.
Here $H_{\rm sat}(T)$ is the field at which the MR becomes
field-independent, i.e., the field-induced AF to FM transition is 
complete. The values of $H_{\rm sat}(T=0)$ are $\simeq$ 3 kOe for sample 4 and 
$\simeq 11.5$ kOe for sample 1.
Figure 10 is very similar to Fig. 1(b) of Ref. [8]
on MBE grown samples for the parallel orientation.

\section{Conclusions}

The temperature dependence of the electrical resistivity and  
magnetoresistance has been studied in ion-beam sputtered Fe-Cr  
multilayers. Typical in-plane negative giant magnetoresistance (GMR) is  
21\% at 10 K saturating at around 1 tesla. Here each Fe layer is  
ferromagnetic  but coupled antiferromagnetically in zero field to the  
neighbouring Fe layers due to the RKKY interaction. This gives rise to  
a high resistance. An external magnetic field aligns the spins in  
different Fe layers producing a ferromagnetic alignment beyond H$_{\rm sat}$ 
which reduces the electrical resistance. The electrical resistivity in  
the fully ferromagnetic state ($H=H_{\rm sat}$) between 5 and 300 K has been  
interpreted as the sum of a residual resistivity, electron-phonon s-d  
scattering and spin-disorder resistivity ( Eq.(5)). The latter has the 
same order of magnitude as in crystalline Fe. 
 
We have calculated the sublattice magnetization $m(T)$ of these Fe-Cr 
multilayers in terms of the planar and interlayer exchange energies 
(Eq.(13) of APPENDIX). The additional spin-dependent scattering in the  
antiferromagnetic state at $H=0$, defined by
$\Delta\rho(T)=\rho(T,H=0)_{AF}-\rho(T,H=H_{\rm sat})_{FM}$,
is obtained from the experimental data 
by assuming that the residual resistivity and the electron-phonon scattering 
are roughly independent of the field. The decrease in $\Delta\rho(T)$ with  
increasing temperature from 5 to 300 K is explained as arising from the  
reduction in the antiferromagnetic order due to the thermal excitation 
of spin waves, i.e., $\Delta\rho(T)\sim m(T)$. Mattson {\em et al.} \cite{two}, on the  
other hand in dc magnetron sputtered Fe-Cr superlattice,
found $\Delta \rho(T) $ decreasing 
as $T^{2}$ at temperature below 100K (Eq. (3)) in contrast to our fits  
(Eq.(7)) over a much wider temperature range.

From our data at intermediate fields $(0<H<H_{\rm sat}$), the spin-dependent part 
of the electrical resistivity,
defined as $\rho_{s}(T)=\rho(T,H)-\rho(T,H>H_{\rm sat})$, 
fits very well to Eq. (4).
We find that $\alpha$ is typically 2 for $H/H_{\rm sat}<1/3$,  
becoming $\sim 1 $ for $H/H_{\rm sat}\simeq2/3$ and then decreasing further
at still higher fields.  
The decrease of the intercept b with increasing H and that of $\Delta\rho(T)$ 
with increasing T (Eq.(7)) are due to 
the decrease of the antiferromagnetic fraction (Eq. (6)) with increasing 
H and T, respectively.
Very similar conclusions were reached by Aliev {\em et al.} \cite{eight} 
in MBE grown Fe-Cr multilayers. 
 
 Finally, we have also obtained the T vs H$_{\rm sat}$ phase diagram for the field-induced AF 
 to FM transition. 

\section{Appendix}

To obtain the magnon energies in the multilayer system,
the following simplified 
Hubbard model is considered on a three-dimensional
lattice consisting of a stack of layers in the z direction
\begin{eqnarray}
H &=& \sum_{{\bf k}\sigma} \epsilon_p ({\bf k}) \;
c_{{\bf k}\sigma}^\dagger c_{{\bf k}\sigma}
-t_z \sum_{i \sigma} c_{i \sigma}^\dagger 
(c_{i+\delta , \sigma} + c_{i-\delta , \sigma} ) \nonumber \\
&+& U\sum_{i} n_{i \uparrow} n_{i \downarrow} \; .
\end{eqnarray}
Here the planar band energy 
$\epsilon_p (k_x,k_y)$ together with the correlation term
describes the ferromagnetic layers, while the 
inter-layer hopping term $t_z$,
which connects sites $i$ to nearest-neighbour sites 
$i\pm \delta $ in the neighbouring layers,
represents the AF exchange coupling between layers. 
We divide the multilayer system into two sublattices with 
alternating A and B layers, 
and consider a ground state in which 
the A-layers have spin polarization 
$\langle n_i ^\uparrow - n_i ^\downarrow \rangle = +m$
in the $+$z direction
while B-layers have spin polarization
$\langle n_i ^\uparrow - n_i ^\downarrow \rangle = -m$     
in the $-$z direction.
The sublattice magnetization $m$, a dimensionless quantity,
measures the AF order parameter
in the multilayer system. 

In this two-sublattice basis,
and in the HF approximation, the Hamiltonian reduces to
\begin{equation}
H = \sum_{{\bf k} \sigma}
\left (
\begin{array}{cc}
a_{{\bf k}\sigma} ^\dagger & b_{{\bf k}\sigma} ^\dagger \\
\end{array}
\right )
\left [
\begin{array}{cc}
\epsilon_p ({\bf k}) -\sigma\Delta & \epsilon_z ({\bf k})  \\
\epsilon_z ({\bf k})& \epsilon_p ({\bf k})+\sigma\Delta  \end{array} 
\right ]
\left (
\begin{array}{c}
a_{{\bf k}\sigma} \\
b_{{\bf k}\sigma} 
\end{array}
\right ) ,
\end{equation}
where $a_{\bf k}$ and $b_{\bf k}$ are the Fourier transforms
of the electronic annihilation operator $c_i$, defined
on the two sublattices A and B, respectively. 
Here $2\Delta = m U$ and the sublattice magnetization $m$ is
determined self-consistently. For simplicity we consider the strong 
correlation limit in which at the HF level $m \approx 1$.
The interlayer band energy $\epsilon_z ({\bf k})=-2t_z \cos k_z$ 
mixes the two ferromagnetic bands,
and hence the quasiparticle band energies  
$E({\bf k})= \epsilon_p ({\bf k})\pm \sqrt {\Delta^2 + 
\epsilon_z ({\bf k})^2}$
have a mixed character with features of both the 
ferromagnetic\cite{ferro}
and antiferromagnetic ground states.\cite{antiferro} 
As the planar (ferromagnetic) band energy $\epsilon_p ({\bf k})$
appears on the diagonal, 
the eigenvectors of the Hamiltonian matrix in Eq. (9) are unchanged
from the AF case.\cite{antiferro}

Evaluation of the magnon propagator $\chi^{-+}({\bf q}\omega)$,
involving transverse spin operators ($S^-$, $S ^+$),
and representing transverse spin fluctuations  
about the Hartree-Fock ordered state,
has been described earlier in the random phase approximation (RPA)
for both ferromagnetic\cite{ferro} and antiferromagnetic\cite{antiferro}
ground states. 
For the multilayer system the magnon propagator is obtained as 
\begin{equation}
\chi^{-+}({\bf q}\omega)= 
\left [
\begin{array}{lr}
J_z + J_p q_p ^2 - \omega  & -J_z \cos q_z  \\
- J_z \cos q_z     & J_z + J_p q_p ^2 + \omega
\end{array}
\right ]
\frac{1}{\omega_{\bf q} ^2 -\omega^2 } \; ,
\end{equation}
for small planar momentum ${\bf q_p}= (q_x,q_y)$.
Here $J_z= 2t_z ^2 /\Delta \approx 4t_z ^2 /U$ is the exchange energy
characterizing the antiferromagnetic coupling between layers,
and $J_p$, the magnitude of which 
depends on details of the planar band energy $\epsilon_p ({\bf k})$,
plays the role of the planar exchange energy.  
The magnon energy $\omega_{\bf q}$ is given by 
\begin{equation}
\omega_{\bf q} ^2 =
(J_p q_p ^2 + J_z ) ^2 - J_z ^2 \cos ^2 q_z \; ,
\end{equation} 
which has the right limiting behaviour 
yielding the antiferromagnetic magnon energy
$J_z \sqrt{1-\cos ^2 q_z }$ as $J_p \rightarrow 0$,
and the ferromagnetic magnon energy 
$J_p q_p ^2 $ as $J_z \rightarrow 0$. 
A completely different starting point in terms of a 
Heisenberg spin model for the multilayer system,
with planar and interlayer exchange energies $J_p$ and $J_z$, 
would yield the same result. 

Going over now to the thermal excitation of magnons in the 
multilayer system, the change $\delta m(T) \equiv  m(T)-m(0) $ 
in the sublattice magnetization at finite temperature $T$
is obtained by considering both the
advanced and retarded modes in the spin-fluctuation propagator
with appropriate Bose weights. After subtracting out the 
zero-temperature (quantum fluctuation) part
the reduction in the sublattice magnetization is obtained as 
\begin{equation}
-\delta m(T) = \int_0 ^ \infty \frac{2\pi q_p dq_p}{(2\pi)^2} 
\int_{-\pi} ^\pi \frac{dq_z}{2\pi} \;
\frac{J_z + J_p q_p ^2 }{\omega_{\bf q}}
\frac{2}{e^{\beta \omega_{\bf q} } -1 } \; .
\end{equation}
Here the upper limit of integration for the $q_p$ integral has been
taken as $\infty$ for convenience, which is valid at 
temperatures low compared to $J_p$, as the high-energy modes
have exponentially small weight. Integration over the planar
momentum $q_p$ finally yields 
\begin{equation}
m(T) = m(0) - \frac{1}{\pi^2} \frac{T}{J_p} 
\int_{0} ^{\frac{\pi}{2}} dq_z \;
\ln \left [ \frac{1}
{1-e^{-\frac{J_z}{T}(1-\cos ^2 q_z )^{1/2}} }
\right ] \; .
\end{equation}

\section*{ACKNOWLEDGMENTS}
One of us (AKM) acknowledges the Physics Department,  
University of Florida, Gainesville 
for local hospitality and experimental facilities.
In addition, discussions with S. B. Arnason and S. Hershfield
are greatfully acknowledged.
This research was supported by a
DOD/AFOSR MURI grant ($\#$ F49620-96-1-0026).

\newpage
\begin{figure}
\caption{Magnetoresistance vs. external field $H$ (kOe) oriented parallel
to the layers for a Xe-ion beam
sputtered Fe-Cr multilayer sample (sample 2) at 10 and 300 K. The MR 
saturates around 13 kOe ($H_{\rm sat}$) and has a typical GMR of 21\% at 10 K.}
\end{figure}

\begin{figure}
\caption{Electrical resistivity ($\rho$) vs. temperature ($T$) for sample
1 at several fields between 0 and 12 kOe. The curves are closer to each
other at higher temperatures indicating that the saturation field
$H_{\rm sat}$ decreases with increasing temperature.}
\end{figure}

\begin{figure}
\caption{Electrical resistivity ($\rho$) vs. temperature (T) data (points)
from 5 to 300 K for samples 1, 2 and 3 at their respective H$_{\rm sat}$. 
The
solid lines are the excellent least-squares fitted curves for fits to 
Eq. (5) which includes lattice and magnetic scattering contributions.}
\end{figure}

\begin{figure}
\caption{The deviation of the actual data from the best fit values is
plotted as a function of temperature ($T$) for sample 2 for fits with and
without the magnetic $T^2$ term. The deviation is much less
($<0.1 \mu\Omega $ cm in 40 $\mu\Omega $cm)
and more random for the fits with the magnetic term than that without it.}
\end{figure}

\begin{figure}
\caption{$\Delta \rho (T) = \rho(T, H=0)_{AF} - 
\rho(T,H=H_{\rm sat})_{FM}$
 vs. temperature($T$) data (stars) for samples 1 and 3. This additional
 spin-dependent resistivity in the AF state decreases with temperature.
 $\Delta\rho$ varies as $T^{2}$ at low temperatures and roughly linearly
 at higher temperatures. The solid lines are the least-squares fitted
 curves for fits to Eq. (7).}
\end{figure}

\begin{figure}
\caption{$\rho_{s}(T)=\rho(T,H)-\rho(T,H_{\rm sat})$ vs. temperature (T)
data (points) for sample 1 at H=0, 3, 4, 6 and 8 kOe. The solid lines 
are the least-squres fitted curves for fits to Eq. (4).}
\end{figure}

\begin{figure}
\caption{$\alpha$ of Eq. (4) vs. H/H$_{\rm sat}$ for all the three samples.
The solid lines are just guides to the eye.}
\end{figure}

\begin{figure}
\caption{$b$ of Eq. (4) vs. AFF(\%) for samples 1 and 3. $b$ 
is found to
increase monotonically with the AFF (both increasing with $H$).}
\end{figure}

\begin{figure}
\caption{Magnetoresistance vs. external field H(kOe) at low fields for 
sample 4(Ar-ion sputtered) for the parallel orientation (H in the film
plane). The temperatures are between 5 K and 205 K (every 20 K) and 300 K.
Clearly the MR $\sim H^2$ at lower fields.}
\end{figure}

\begin{figure}
\caption{ $T$ vs. $H_{\rm sat}$ phase diagram for samples 4
(Ar-ion sputtered) and 1 (Xe-ion sputtered).
Here $H_{\rm sat}$ is the field at which the MR
becomes field independent, i.e., the field-induced AF to FM transition
is complete. The values of $H_{\rm sat}(T=0)$ are 3 kOe for sample 4 and
11.5 kOe for sample 1.}

\end{figure}

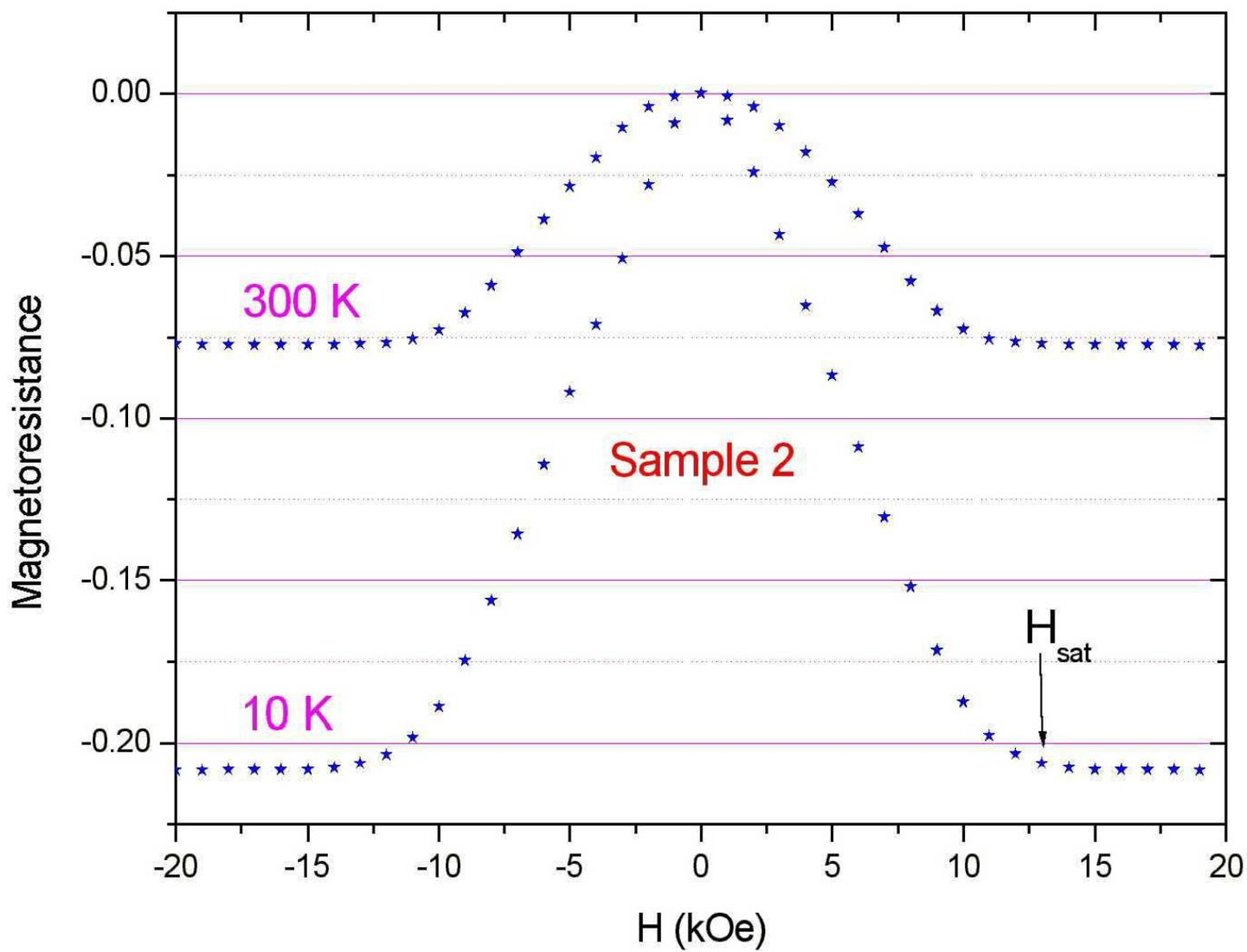
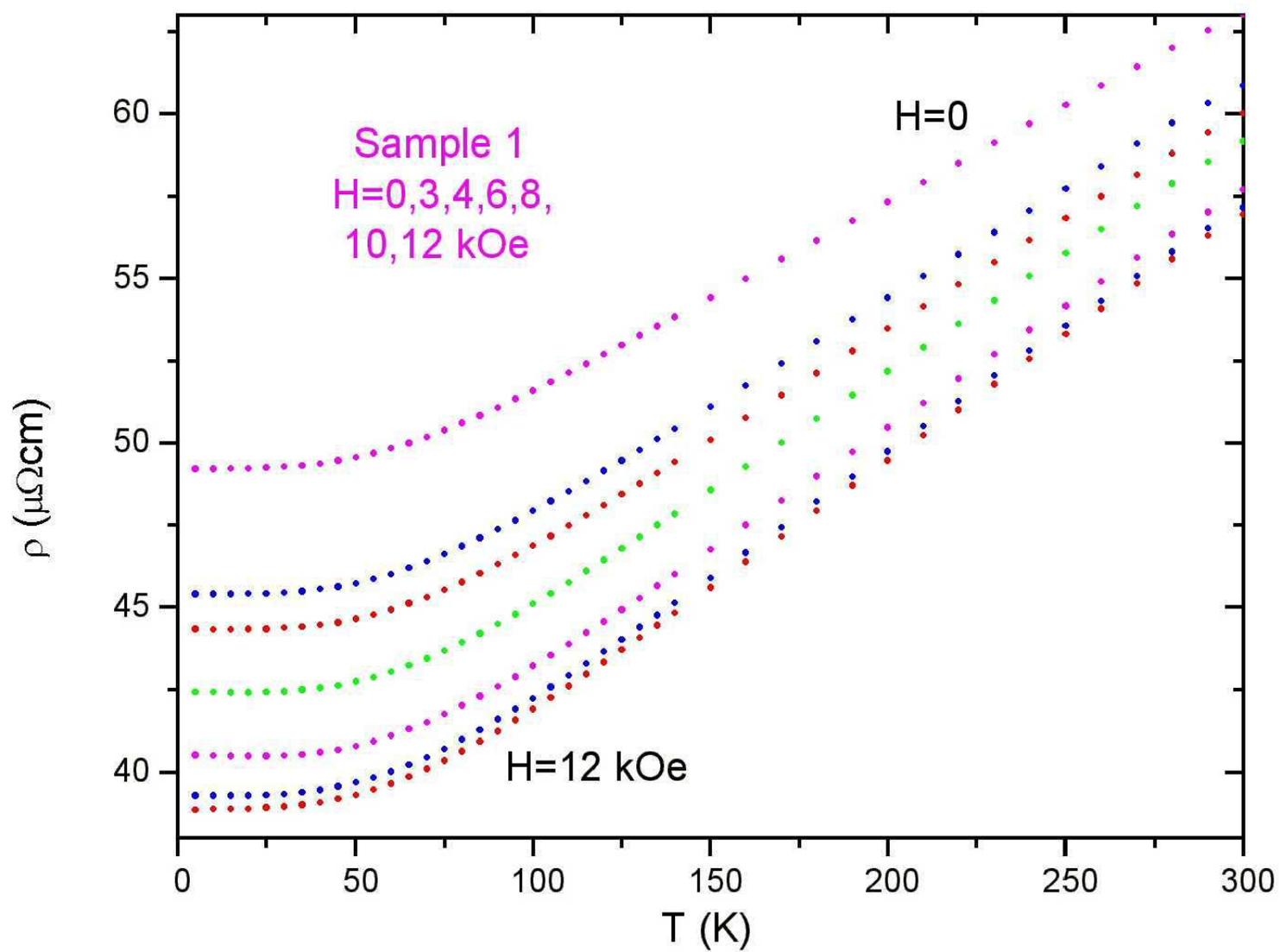
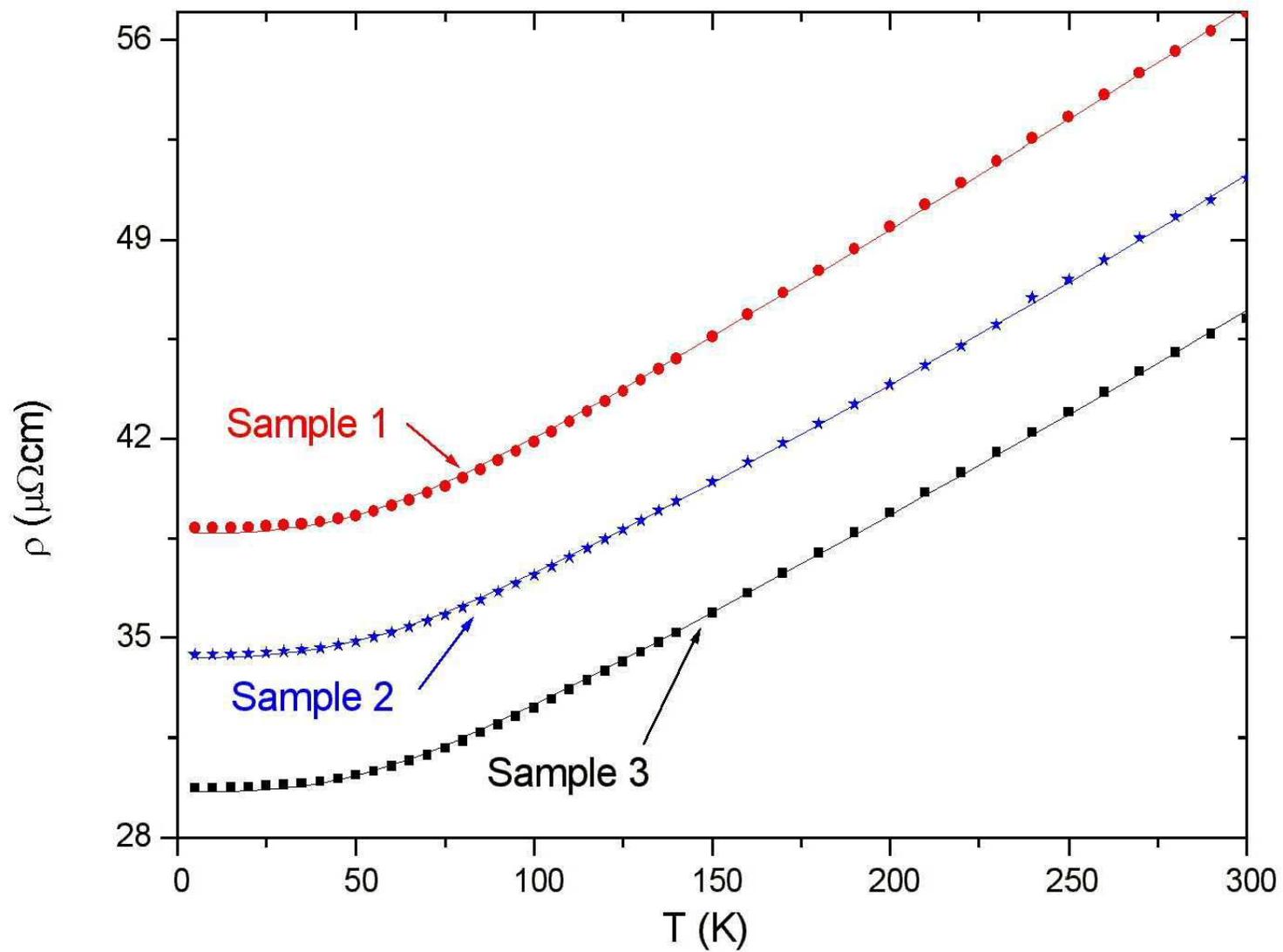
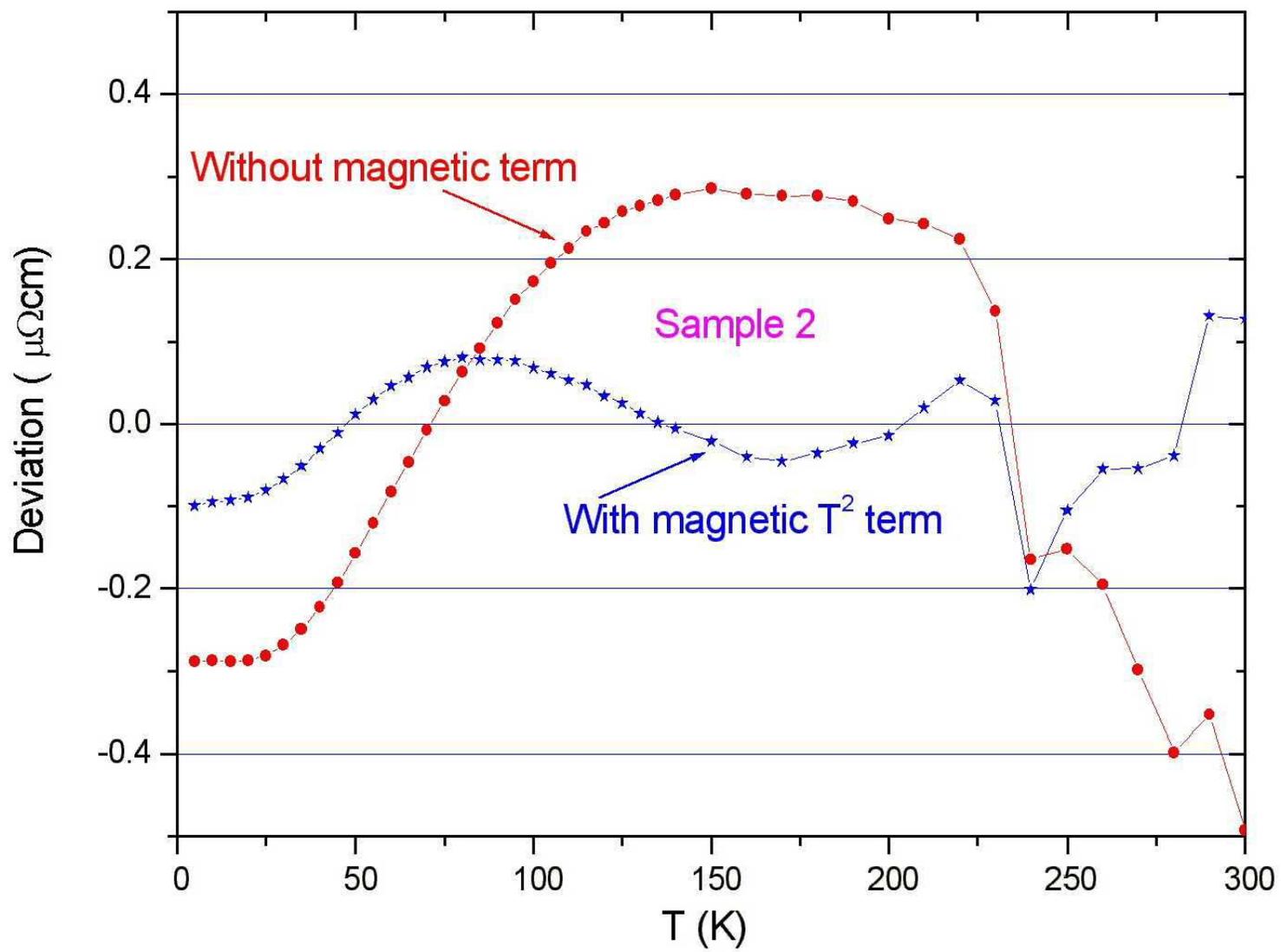
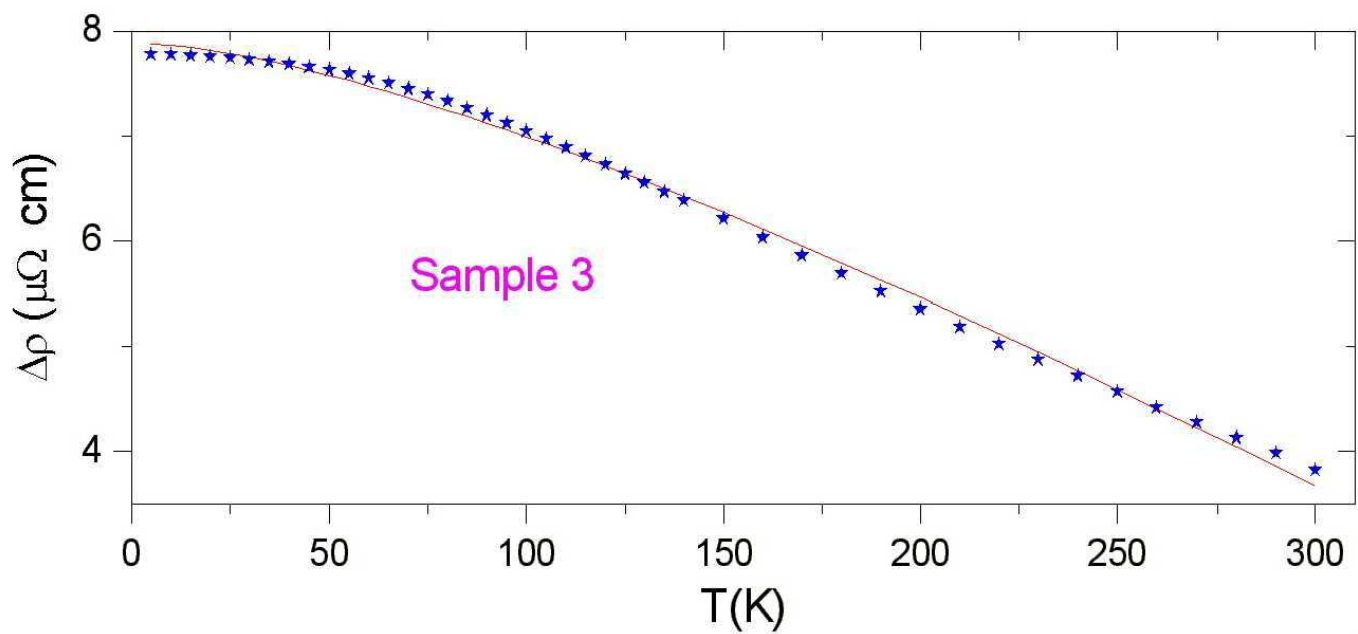
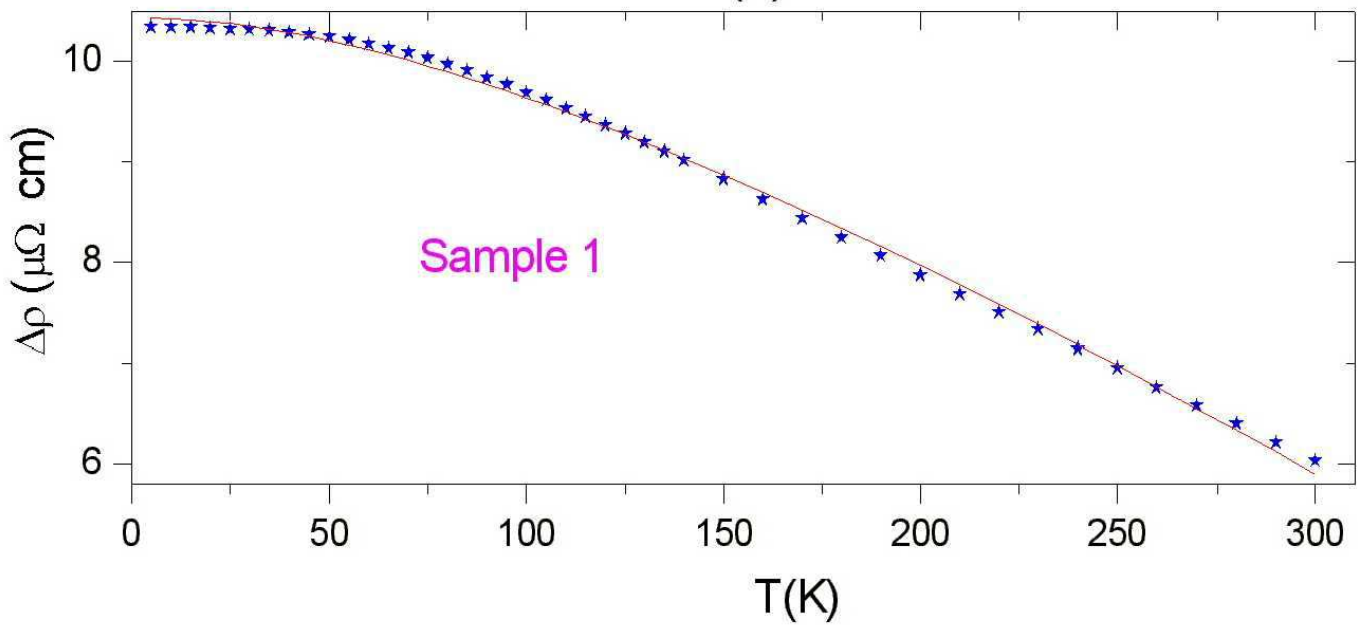
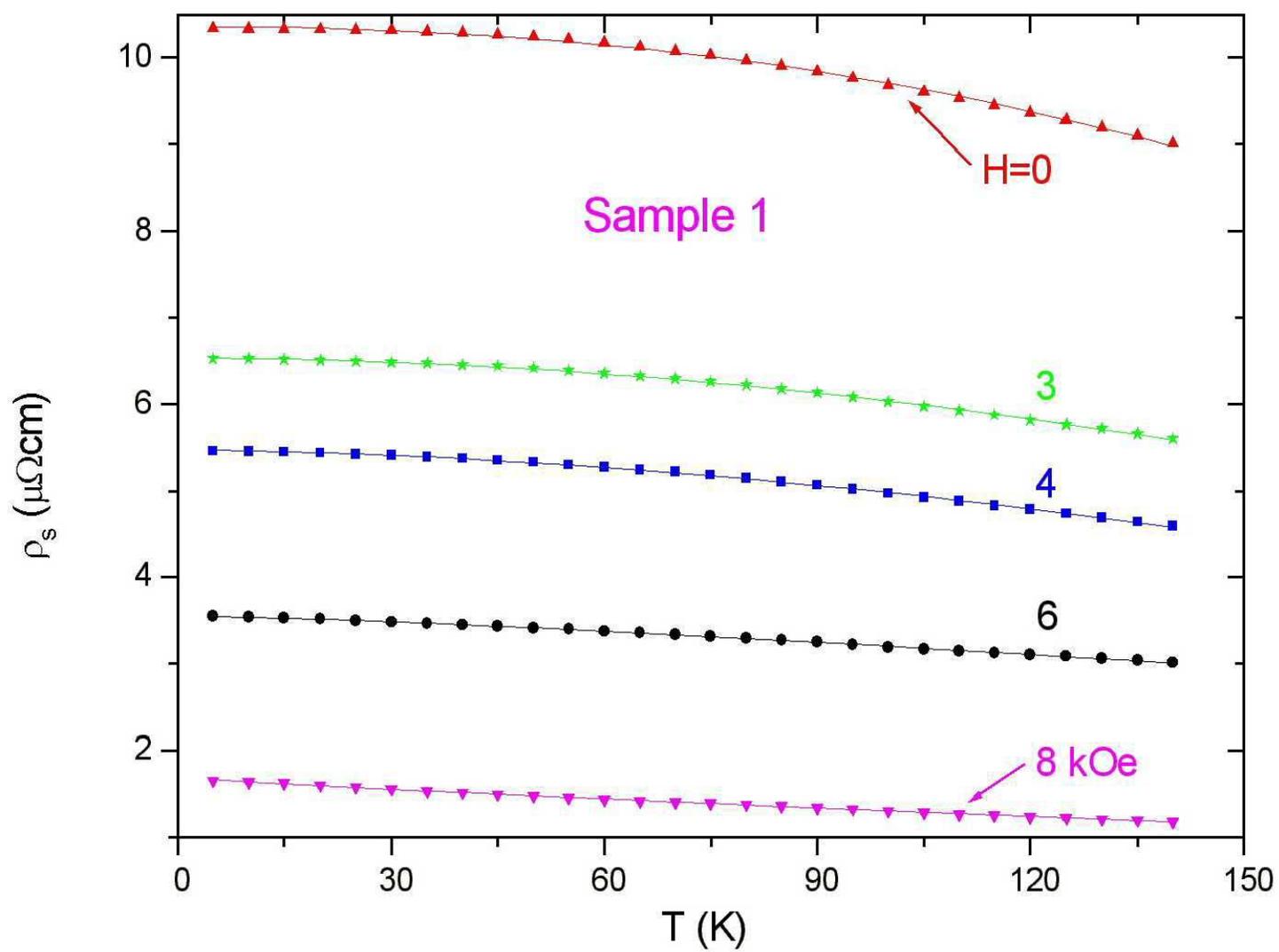
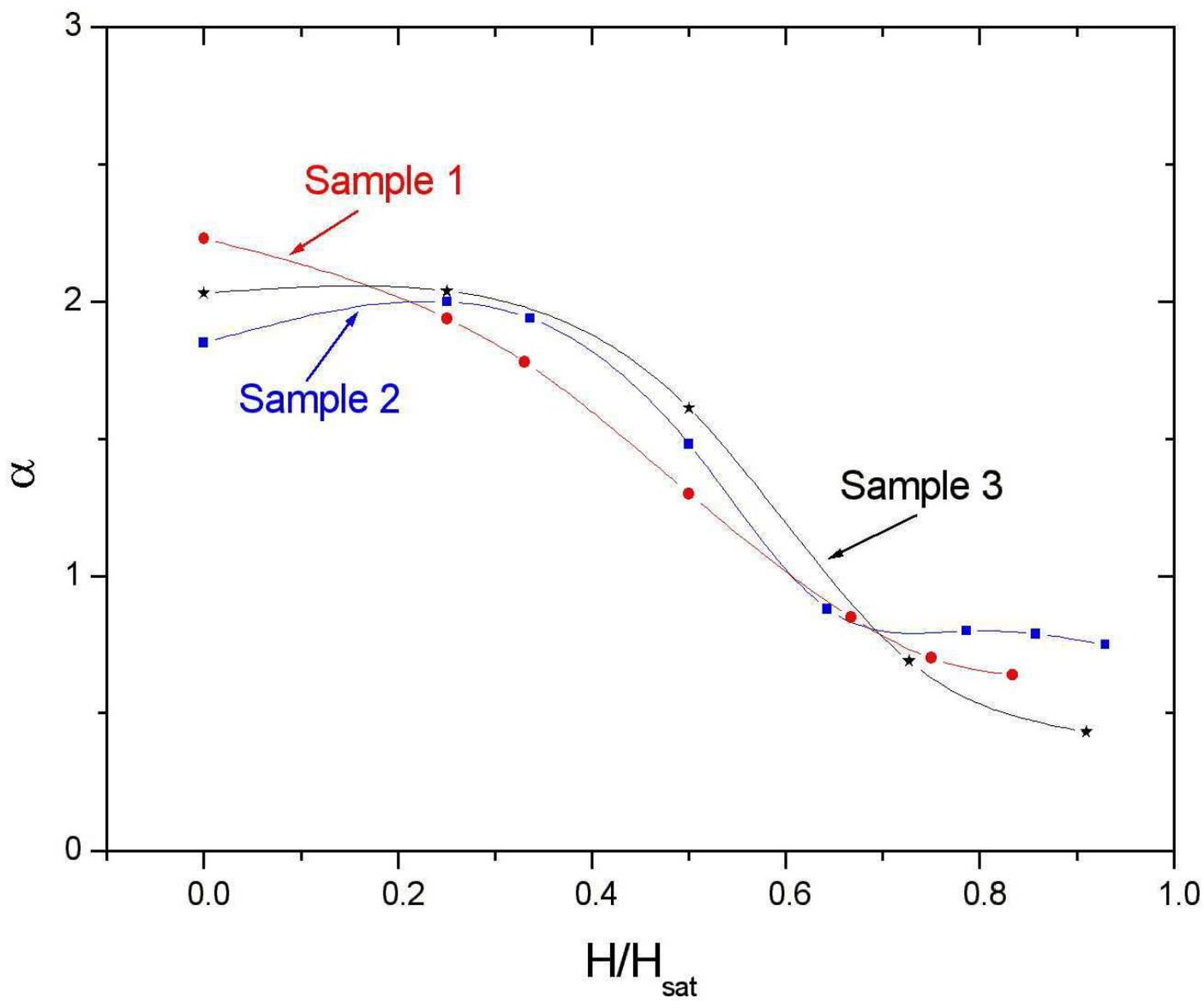
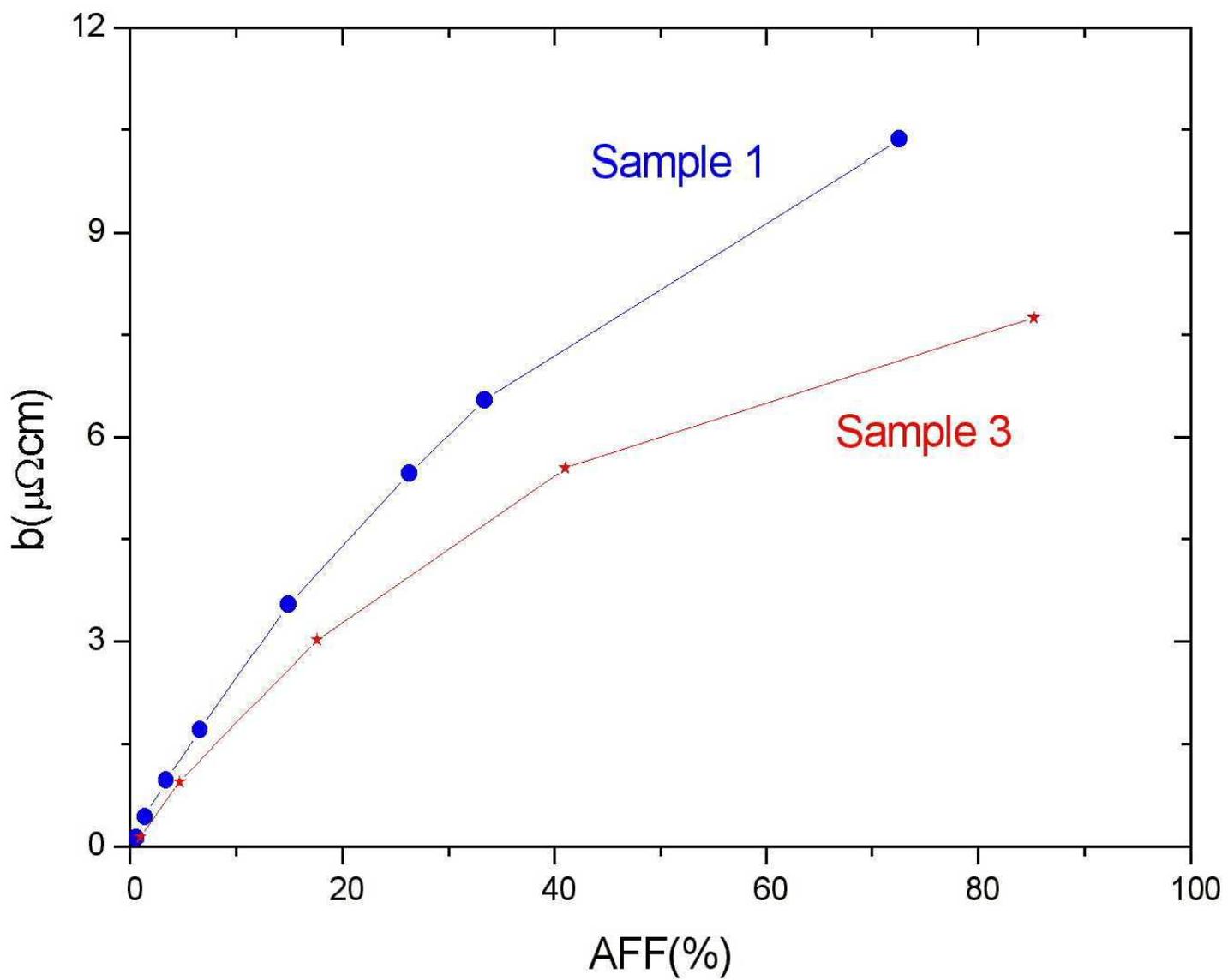
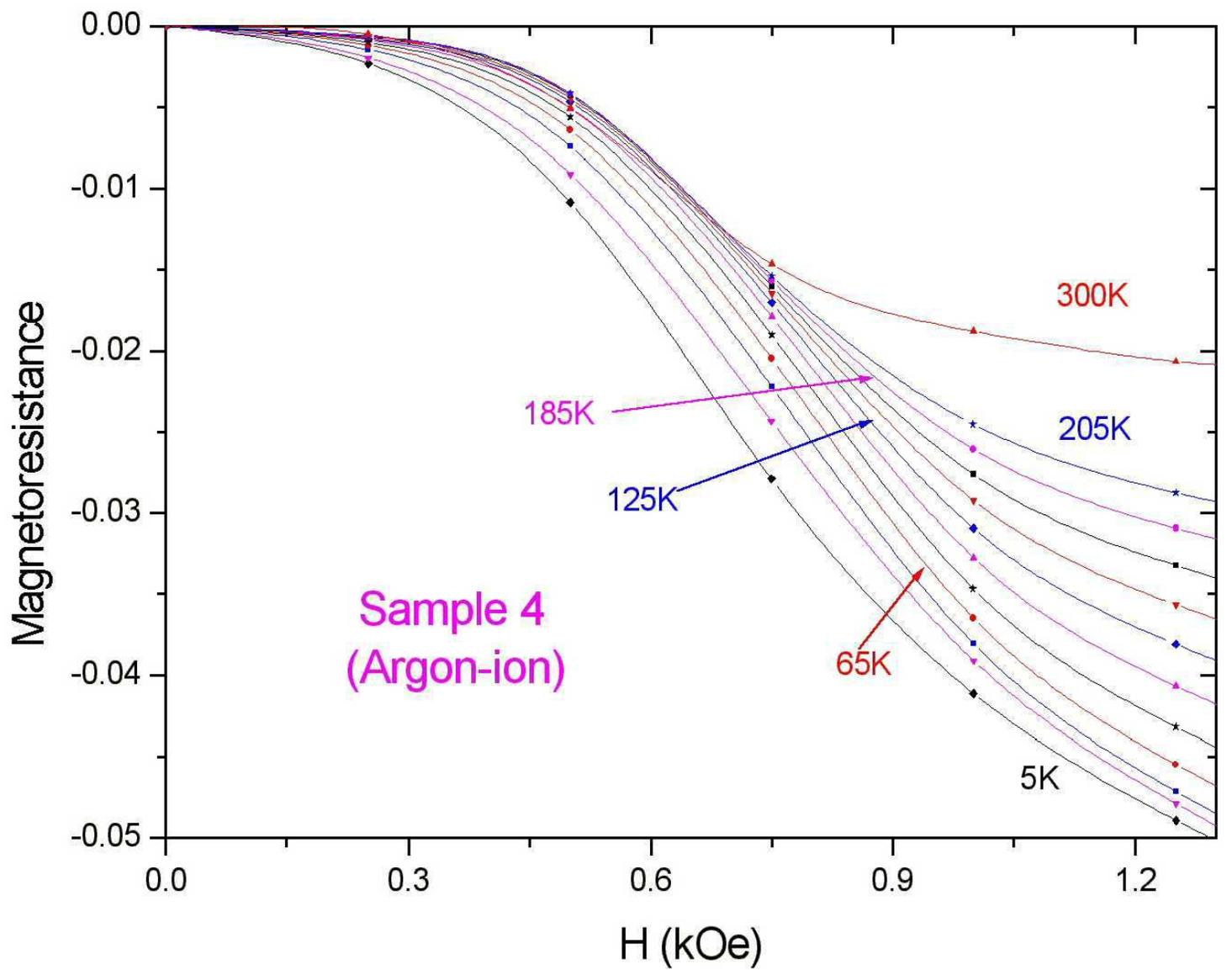
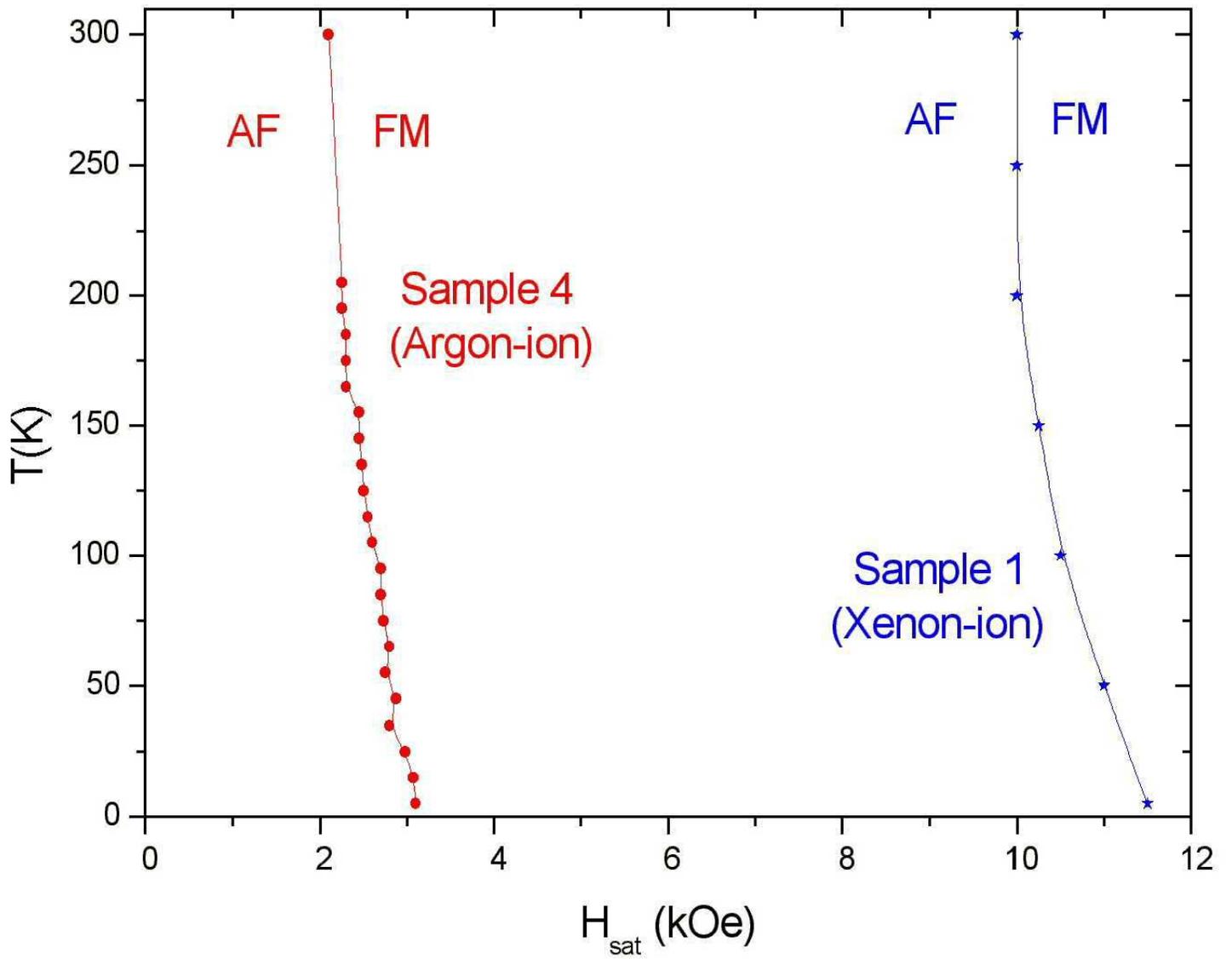
\end{document}